\documentclass[reqno,10pt]{amsart}
\usepackage{amssymb,amsmath,latexsym,amsthm}
\usepackage{color,mathrsfs}
\usepackage{url}
\usepackage{enumerate}
\usepackage[square,sort,comma,numbers]{natbib}
\usepackage{dsfont}
\usepackage{diagbox}
\usepackage{graphicx}
\usepackage{colortbl}
\usepackage{float}
\usepackage{color}
\usepackage[colorlinks=true, linkcolor=blue_links,
urlcolor=blue_links, citecolor=blue_links]{hyperref}
\definecolor{blue_links}{RGB}{13,0,180} 
\definecolor{lightblue}{rgb}{0.9,0.9,1}

\usepackage{endnotes}

\renewcommand{\textbf}[1]{\begingroup\bfseries\mathversion{bold}#1\endgroup}

\usepackage{fancyhdr}
\usepackage{pstricks,pst-plot,pst-node,pstricks-add}

\newtheorem{thm}{Theorem}[section]

\newtheorem{Conjecture}[thm]{Conjecture}

\newcommand{\argmin}{\mathop{\rm argmin}\nolimits}

\definecolor{orange-red}{rgb}{1.0, 0.27, 0.0}

\newcommand{\R}{\mathbb R}

\newcommand{\Z}{\mathbb Z}
\newcommand{\N}{\mathbb N}
\newcommand{\C}{\mathbb C}
\numberwithin{equation}{section}

\def\XXint#1#2#3{{\setbox0=\hbox{$#1{#2#3}{\int}$}
    \vcenter{\hbox{$#2#3$}}\kern-.5\wd0}}

\allowdisplaybreaks

\begin{document}
\title{Three-dimensional lattice ground states for Riesz and Lennard-Jones type energies}

\author{Laurent B\'{e}termin}
\address[Laurent B\'{e}termin]{Institut Camille Jordan, Universit\'e Claude Bernard Lyon 1, 69622 Villeurbanne, France}
\email{betermin@math.univ-lyon1.fr}
 
 \author{Ladislav \v{S}amaj}
\address[Ladislav \v{S}amaj]{Institute of Physics, Slovak Academy of Sciences, 
84511 Bratislava, Slovakia}
\email{fyzimaes@savba.sk}

 \author{Igor Trav\v{e}nec}
\address[Igor Trav\v{e}nec]{Institute of Physics, Slovak Academy of Sciences,
 84511 Bratislava, Slovakia}
\email{fyzitrav@savba.sk}

\begin{abstract}
The Riesz potential $f_s(r)=r^{-s}$ is known to be an important building block
of many interactions, including Lennard-Jones type potentials
$f_{n,m}^{\rm{LJ}}(r):=a r^{-n}-b r^{-m}$, $n>m$ that are widely used in
Molecular Simulations. In this paper, we investigate analytically and
numerically the minimizers among three-dimensional lattices of Riesz and
Lennard-Jones energies. We discuss the minimality of the Body-Centred-Cubic
lattice (BCC), Face-Centred-Cubic lattice (FCC), Simple Hexagonal lattices
(SH) and Hexagonal Close-Packing structure (HCP), globally and at fixed
density. In the Riesz case, new evidence of the global minimality at fixed
density of the BCC lattice is shown for $s<0$ and the HCP lattice is computed
to have higher energy than the FCC (for $s>3/2$) and BCC (for $s<3/2$)
lattices. In the Lennard-Jones case with exponents $3<m<n$,
the ground state among lattices is confirmed to be a FCC lattice whereas
a HCP phase occurs once added to the investigated structures.
Furthermore, phase transitions of type ``FCC-SH" and ``FCC-HCP-SH"
(when the HCP lattice is added) as the inverse density $V$ increases
are observed for a large spectrum of exponents $(n,m)$. In the SH phase,
the variation of the ratio $\Delta$ between the inter-layer distance $d$ and
the lattice parameter $a$ is studied as $V$ increases.
In the critical region of exponents $0<m<n<3$,
the SH phase with an extreme value of the anisotropy parameter $\Delta$ dominates. 
If one limits oneself to rigid lattices, the BCC-FCC-HCP phase diagram is found.
For $-2<m<n<0$, the BCC lattice is the only energy minimizer.
Choosing $-4<m<n<-2$, the FCC and SH latices become minimizers.
\end{abstract}

\subjclass[2010]{74G65, 70G75, 74N05.}

\keywords{Riesz potential, Lennard-Jones potential, Lattices, Ground states, Epstein zeta function.}

\maketitle

\section{Introduction, setting and main results}

\subsection{Motivation}
From the first attempt by Huygens \cite{Huygens} to describe solids as periodic assemblies of objects to the formalization of the Crystallization Conjecture by Blanc and Lewin \cite{BlancLewin-2015}, the mathematical journey to rigorously justify the emergence of periodic structures in crystal solids never stopped to attract new travelers. Nevertheless, only few rigorous results showing the global minimality, i.e. among all possible configurations, of lattice structures for interaction energies are available. Beside the one-dimensional case \cite{VN2,BlancLebris,CohnKumar,SandierSerfaty1d,Crystbinary1d} where the optimality of the equidistant configuration $\Z$ is easily derivable, the only known results in dimension $d=2$ \cite{Rad2,Crystal,Suto1,ELi,Stef1,Stef2,BDLPSquare,DelucaFriesecke-2018} and $d=3$ \cite{Suto1,TheilFlatley} rely on perturbations of hard-sphere potentials and specific angular dependence.

In dimension $d\in \{8,24\}$, the linear programming method initiated by Cohn an Elkies \cite{CohnElkies} has lead to recent important results for packings \cite{Viazovska,CKMRV} and crystallization at fixed density \cite{CKMRV2Theta,PetracheSerfatyCryst,BDefects20}. In particular they showed the universal optimality, at any fixed density and for pairwise energies with interaction potentials $f(r)=F(r^2)$ where $F$ is a completely monotone function, of the densest packings in these dimensions -- namely the Gosset lattice $\mathsf{E}_8$ and the Leech lattice $\Lambda_{24}$. These are the only crystallization results that apply at any fixed density to Riesz energies, where the interaction potential is
$$
f_s(r)=\frac{1}{r^s},
$$ 
and at sufficiently high fixed density for Lennard-Jones type energies (see \cite[Thm. 2.17]{BDefects20}) for which the interaction potential is defined by
$$
f_{n,m}^{\rm{LJ}}(r)=\frac{a}{r^n}-\frac{b}{r^m},
$$
where $(a,b)\in (0,\infty)$ and $n>m>d$. The same minimality property is conjectured by Cohn and Kumar \cite{CohnKumar} to hold in dimension $d=2$ for the triangular lattice $\mathsf{A}_2$ whereas it is known that no universal minimizer exists in dimension $d=3$ \cite{SarStromb}. 

The potentials studied in this paper are of high importance in Mathematics and Physics, and their lattice ground states are known to be fundamental in many domains. On the one hand, Riesz type energies arise in Mathematical Physics and Number Theory in many ways. For instance, in the Coulomb case $s=d-2$ (or $f(r)=-\log r$ in dimension $d=2$), the Wigner Conjecture for Jellium \cite{WignerConj} states that electrons embedded in an uniform background of positive charges must crystallize on a triangular and a Body-Centred-Cubic lattice in dimensions 2 and 3 respectively (see also \cite{BlancLewin-2015,LewinJelliumReview}). This is also called the Abrikosov Conjecture \cite{Abrikosov} or Vortices Conjecture \cite{Sandier_Serfaty} in the two-dimensional setting related to the vortices in the Ginzburg-Landau theory of superconductors of type II. An equivalent problem can be stated in higher dimension and for general Riesz interaction \cite{SerfRoug15,JMJ:9726774,PetracheSerfatyCryst,PetracheCotar17,FloatingLewin,Lauritsen,LewinJelliumReview}. The general Riesz energy and its related minimization problem also appear in the theory of Random Point Configurations \cite{GhoshRandom} as well as Approximation Theory \cite{Sobolev} and Number Theory \cite{OPS,SarStromb,CoulLazzarini}. On the other hand, Lennard-Jones potentials have been introduced by Mie \cite{Mie} and popularized by Jones \cite{LJ} in its classical $(n,m)=(12,6)$ form, i.e. when the interaction is of Van der Waals type $\sim \hspace{-1mm} b r^{-6}$ for large $r$, initially for studying gas argon. The repulsion at short distance of type $\sim a r^{-n}$ is here to mimic Pauli exclusion Principle. Since $f_{n,m}^{\rm{LJ}}$ is a bonding potential, i.e. it has an equilibrium distance, it has been widely used in Molecular Simulation (see e.g. \cite{MolecSimul,Kaplan,LJWang}). It also appears to be a good model for social aggregation \cite{MEKBS} and has many other applications, like in Robotics \cite{LJRobot}. In dimension $d=2$, numerical investigations suggest that the ground state of the classical Lennard-Jones energy (i.e. $(n,m)=(12,6)$) is a triangular lattice \cite{YBLB}, whereas the Hexagonal Close-Packing structure is expected to be the one in dimension $d=3$ \cite{ModifMorse}. Notice also that the ground states of three-dimensional finite-range Lennard-Jones energy among a finite number of structures has been also studied in \cite{LJOrtner} for different truncation, cutoff distances, pressure and exponents. Notice also that the lattice ground states for Lennard-Jones type energies have been recently proven to be related to the one of the Embedded-Atom Models with Riesz-type electron density and Riesz or Lennard-Jones nuclei interaction, see \cite{LBEAM21}.

\subsection{Minimization among lattices and setting}
The goal of this paper is to investigate the possible lattice ground states for Riesz and Lennard-Jones energies in dimension $d=3$, but in the case where simple periodicity is assumed. Indeed, once restricted to the class of lattices 
$$
\mathcal{L}_d:=\left\{L= \bigoplus_{i=1}^d \Z u_i\subset \R^d :\textnormal{$\{u_i\}$ is a basis of $\R^d$}  \right\},
$$
the above minimization problem becomes simpler and the Riesz and Lennard-Jones type lattice energies are respectively defined by 
$$
\zeta_L(s):=\frac{1}{2}\sideset{}{'}\sum_{p\in L} \frac{1}{|p|^s},\quad \textnormal{and}\quad E_{n,m}^{\rm{LJ}}[L]:=\frac{1}{2}\sideset{}{'}\sum_{p\in L}f_{n,m}^{\rm{LJ}}(|p|)=a\zeta_L(n)- b\zeta_L(m),
$$
where $\zeta_L(s)$ is the Epstein zeta function \cite{Epstein1} associated to the lattice $L$ and $\sideset{}{'}\sum$ means that we sum on all the points excepted the origin. Notice that a factor $1/2$ has been added in order to identify the Epstein zeta function with the energy per point of $L$ with interaction potential $f_s(r)=r^{-s}$. Recall that $s\mapsto \zeta_L(s)$ admits an analytic continuation on $\C \backslash\{d\}$ (see also \cite{EliRom}) and can therefore be defined for any $s<d$. In particular, the value of $\zeta_L(s)$ when $s\in (d-4,d)$ is the Jellium energy of the lattice $L$ for general Riesz interaction \cite[Thm 3.1]{Lauritsen} including the Coulomb case $s=d-2$, originally proved by Cotar and Petrache for $s\in [d-2,d)$ in \cite[Lemma 2.6]{PetracheCotar17} (see also \cite{FloatingLewin}). Notice that the equivalence between analytic continuation and renormalized energy (of Jellium type) has been recently investigated in a broader framework by Lewin in its review on  Riesz and Coulomb gases \cite[Section IV]{LewinJelliumReview}.

Whereas in dimension $d=2$, the triangular lattice has been shown to be minimal for $L\mapsto \zeta_L(s)$ for all $s>0$ in the set of lattices 
$$
\mathcal{L}_d(V):=\{ L\in \mathcal{L}_d : |\det(u_1,...,u_d)|=V\}\subset \mathcal{L}_d,
$$
for any fixed covolume $V>0$ (also called ``inverse density"), see \cite{Rankin,Eno2,Cassels,Diananda,Mont}, no such optimality result is shown in dimension $d=3$. Only conjectures \cite{SarStromb} and local minimality results \cite{Ennola,DeloneRysh,Gruber,LBbonds21} are available (see Section \ref{sec-Riesz} for more details).

The same holds for the Lennard-Jones type energy $E_{n,m}^{\rm{LJ}}$. Indeed, the two-dimensional ground state of $E_{n,m}^{\rm{LJ}}$ in $\mathcal{L}_2$ has been proven to be a triangular lattice for many exponents $(n,m)$ by the first author \cite{BetTheta15,LBLJComput2021}. Furthermore, we have observed and partially proved a phase transition of type ``triangular-rhombic-square-rectangular" for the minimizer of $E_{n,m}^{\rm{LJ}}$ in $\mathcal{L}_2(V)$ as $V$ increases, see \cite{Beterloc,SamajTravenecLJ}. Only asymptotic and  local minimality results have been shown in dimension $d=3$ in \cite{Beterminlocal3d,OptinonCM,LBbonds21} (see Section \ref{sec-LJ} for more details). Notice that similar problems have been investigated by the first author in \cite{BFK,BFMaxTheta20,BFS} concerning charged systems and related lattice optimality for frame bounds in Time-Frequency Analysis.

We aim to present a complete picture of the lattice ground states of $L\mapsto \zeta_L(s)$ and $E_{n,m}^{\rm{LJ}}$ in $\mathcal{L}_3$ and $\mathcal{L}_3(V)$. In particular, we want to show minimality properties of the following important three-dimensional lattices (given here with unit density):
\begin{align}
\textnormal{The Simple Cubic lattice (SC)} \quad &\Z^3:=\Z(1,0,0)\oplus \Z(0,1,0)\oplus \Z(0,0,1)\\
\textnormal{The Face-Centred-Cubic lattice (FCC)}\quad &\mathsf{D}_3:=2^{-\frac{1}{3}}\left[\Z(1,0,1)\oplus \Z(0,1,1)\oplus \Z(1,1,0)  \right]\\
\textnormal{The Body-Centred-Cubic lattice (BCC)}\quad &\mathsf{D}_3^*:=2^{\frac{1}{3}}\left[\Z(1,0,0)\oplus \Z(0,1,0)\oplus \Z\left(\frac{1}{2},\frac{1}{2},\frac{1}{2}  \right)  \right]\\
\textnormal{The Simple Hexagonal lattice (SH)}\quad &\mathsf{A_f}(\Delta):=\left( \frac{2}{\sqrt{3}\Delta}\right)^{\frac{1}{3}}\left[\Z(1,0,0)\oplus\Z\left( \frac{1}{2},\frac{\sqrt{3}}{2},0\right)\oplus \Z(0,0,\Delta)\right].
\end{align}
Notice that the SH lattices are non-shifted stacking of triangular lattices with lattice constant $a>0$ and inter-layer distance $d>0$ and we set $\Delta=d/a$. Furthermore, we also want to investigate the particular role of the unit density hexagonal close-packed structure, which is not a lattice as defined above, defined by
\begin{align}
\textnormal{The Hexagonal Close-Packed structure (HCP)}\quad &\mathsf{A3}:= \Lambda\cup \left( \Lambda +\left(\frac{1}{2},\frac{1}{\sqrt{12}},\sqrt{\frac{2}{3}}  \right) \right)\\
& \Lambda:=\Z(1,0,0)\oplus \Z\left(\frac{1}{2},\frac{\sqrt{3}}{2},0 \right)\oplus \Z\left( 0,0,\sqrt{\frac{8}{3}} \right).
\end{align}

\begin{figure}[!h]
\includegraphics[clip,width=3cm]{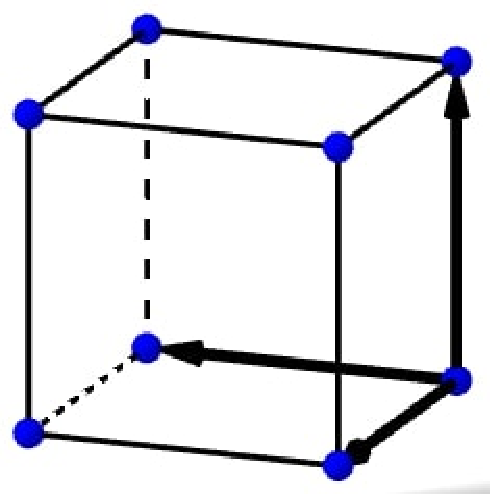} \quad \includegraphics[clip,width=3cm]{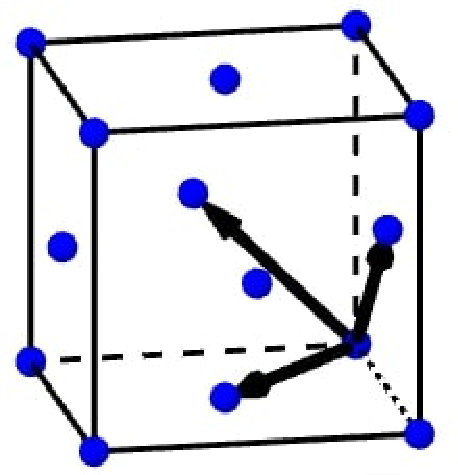}\quad
\includegraphics[clip,width=3cm]{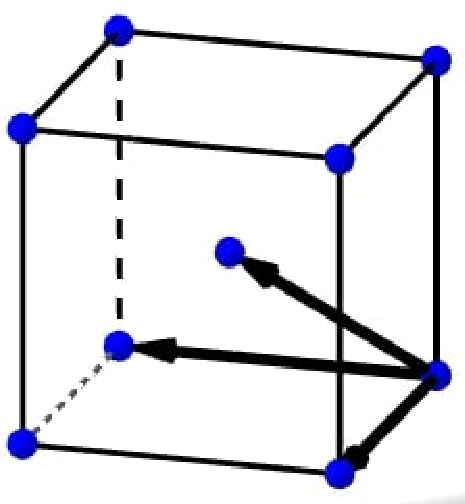}\\
\includegraphics[clip,width=47mm]{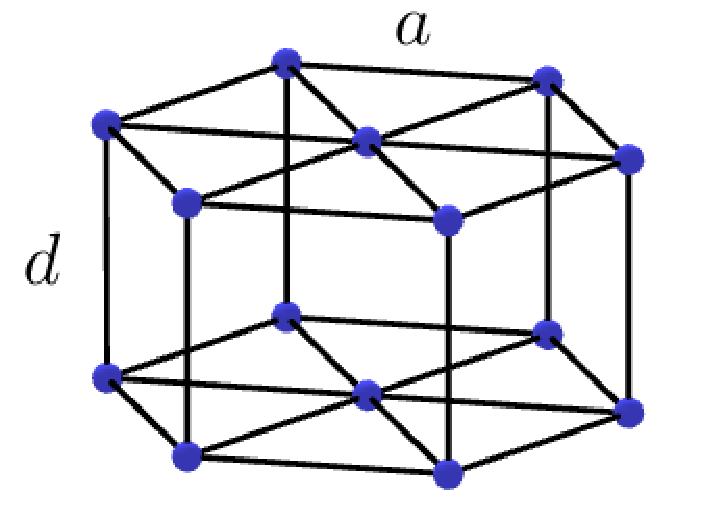} \quad
\includegraphics[clip,width=4cm]{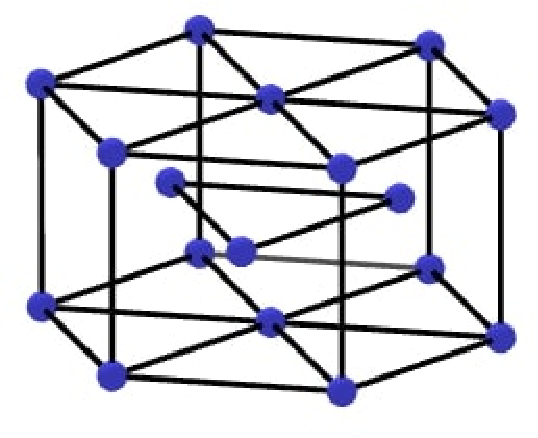}
\caption{Main configurations of $\widetilde{\mathcal{L}_3}$: the SC lattice $\Z^3$ (up left), the FCC lattice $\mathsf{D}_3$ (up middle), the BCC lattice $\mathsf{D}_3^*$ (up right), the SH lattice $\mathsf{A_f}(\Delta)$ (down left), the anisotropy parameter $\Delta=d/a$, and the HCP lattice $\mathsf{A3}$ (down right).}
\label{fig-lattices}
\end{figure}

All the above mentioned periodic structures are depicted in Figure \ref{fig-lattices}. Notice that the names chosen for the SH and HCP structures are the one given by the \textit{Strukturbericht designation} \cite{Aflow}. Furthermore, we define for convenience the new sets, i.e. the sets of lattices with added HCP structure,
\begin{align}
\widetilde{\mathcal{L}_3}(V):=\mathcal{L}_3(V)\cup V^{\frac{1}{3}}\mathsf{A3}, \quad \textnormal{and}\quad \widetilde{\mathcal{L}_3}:=\bigcup_{V>0} \widetilde{\mathcal{L}_3}(V).
\end{align}

\subsection{Method}

Whereas different rigorous techniques exist for studying lattice energies in dimension $d=2$ (see e.g. \cite{Mont,BetTheta15,BDefects20,BFMaxTheta20,BFS}), where $\dim \mathcal{L}_2=3$ and $\dim \mathcal{L}_2(V)=2$ for all $V>0$, the situation in dimension $d=3$ is much more complicated. Indeed the so-called fundamental domain containing only one copy of each lattice (up to dilation, isometry and symmetry) is $5$-dimensional for $\mathcal{L}_3(1)$ and has a complicated shape (see e.g. Terras \cite{Terras_1988}). Very few rigorous results have been proven in dimension 3. Let us recall as an example the computer-assisted proof of Sarnak and Str\"ombergsson \cite{SarStromb} where the FCC lattice is shown to be minimal for the height of the flat torus (i.e. the derivative of $s\mapsto \zeta_L(s)$ at $s=0$).

In our work, we parametrize any lattice $L\in \mathcal{L}_3$ by $6$ real numbers $(u,v,x,y,z,V)$ where $V$ is the covolume of $L$ and, following \cite{Beterminlocal3d},
\begin{align*}
&L:=V^{\frac{1}{3}}\left[\Z u_1\oplus \Z u_2\oplus \Z u_3\right], \\
& u_1=2^{\frac{1}{6}}\left(\frac{1}{\sqrt{u}},0,0  \right),\quad u_2=2^{\frac{1}{6}}\left(\frac{x}{\sqrt{u}},\frac{v}{\sqrt{u}},0  \right),\quad u_3=2^{\frac{1}{6}}\left(\frac{y}{\sqrt{u}},\frac{vz}{\sqrt{u}}, \frac{u}{v\sqrt{2}}  \right).
\end{align*}
This parametrization appears to be the same as in \cite[Eq. (27)]{SarStromb} where $(t_{12},t_{13}, t_{23})=(x,y,z)$, $y_1=v^{-2}$ and $y_2=2 u^{-3} v^4$. We therefore use the corresponding three-dimensional Grenier's fundamental domain shifted by one half in order to have only one copy of the unit density FCC point in it, i.e. $(u,v,x,y,z)\in \mathcal{G}_3$ defined by the following 9 inequalities:
\begin{enumerate}
\item[(i)] $1\leq (1+x-y)^2 + v^2\left((1-z)^2 +\frac{u^3}{2v^4} \right)$
\item[(ii)] $1\leq (x-y)^2 + v^2\left((1-z)^2 +\frac{u^3}{2v^4} \right)$
\item[(iii)] $1\leq x^2 +v^2$
\item[(iv)] $1\leq y^2 + v^2\left(z^2+  \frac{u^3}{2v^4}\right)$
\item[(iv')] $1\leq (y-1)^2 + v^2\left(z^2+  \frac{u^3}{2v^4}\right)$
\item[(v)] $1\leq z^2 + \frac{u^3}{2v^4}$
\item[(vi)-(viii)] $0\leq x\leq \frac{1}{2}$, $0\leq y\leq 1$, $0\leq z\leq \frac{1}{2}$.
\end{enumerate}
Our goal is then to investigate the minimizers of the following type of energy per point
$$
E_f[L]:=\frac{1}{2}\sideset{}{'}\sum_{p\in L} f(|p|^2)=\frac{1}{2}\sideset{}{'}\sum_{j,k,l}f\left(\frac{2^{\frac{1}{3}} V^{\frac{2}{3}}}{u}\left[ (j+xk+yl)^2+v^2(k+zl)^2+\frac{u^3}{2v^2}l^2 \right]  \right),
$$
when $(u,v,x,y,z)\in \mathcal{G}_3$, $V>0$ and $f(r)\in \{ r^{-\frac{s}{2}}, a r^{-\frac{n}{2}}-br^{-\frac{m}{2}}\}$. Notice that when our potential $f$ is not summable (i.e. when $m,n,s\leq 3$), we have used the analytic continuation of the Epstein zeta function (see e.g. \cite{EliRom}), but the corresponding expression can always be written in terms of energies of type $E_f$.

To compute numerically the minimizers, we have used the optimization tool \texttt{FindMinimum} in Mathematica, which includes \texttt{ConjugateGradient}, \texttt{PrincipalAxis}, \texttt{LevenbergMarquardt}, \texttt{Newton}, \texttt{QuasiNewton}, \texttt{InteriorPoint}, and \texttt{LinearProgramming}  methods. It has to be noticed that, as in \cite{SarStromb}, we can systematically restrict our study to a compact domain of $\mathcal{G}_3$ since our energy diverges or goes to zero as the parameters $(u,v)$ go to infinity (see also \cite{LBLJComput2021} for such example in two dimensions).

When $L\in \{\Z^3, \mathsf{D}_3, \mathsf{D}_3^*, \mathsf{A_f}(\Delta),\mathsf{A3}  \}$, the exact formulas of $E_f[L]$ and their analytic continuations can be found in Section \ref{sec-formulas}. Also, since the shape (i.e. its class up to dilation and isometry) of the global minimizer of $E_{n,m}^{\rm{LJ}}$ is independent of $(a,b)$ (see \cite{OptinonCM}), we have decided to choose $a=\frac{m}{n-m}$ and $b=\frac{n}{n-m}$ as it is usually done in the Physics literature. Therefore, our Lennard-Jones type energy will be, for any lattice $L$,
\begin{equation}\label{elj}
E^{\rm LJ}_{n,m}[L]=\frac{m}{n-m}\zeta_L(n)-\frac{n}{n-m}\zeta_L(m).
\end{equation}

\textbf{Plan of the paper.} Section \ref{sec-Riesz} and \ref{sec-LJ} are respectively devoted to investigate the lattice ground states for Riesz and Lennard-Jones type energies. The exact formulas used to compute our energies are stated in Section \ref{sec-formulas}.

\section{Minimizers of the Riesz energy}\label{sec-Riesz}

In this part, we investigate the minimizers of the Epstein zeta function $L\mapsto \zeta_L(s)$ in $\mathcal{L}_3(1)$ or $\widetilde{\mathcal{L}_3}(1)$ and for any $s\in \R$. The formulas we are using are available in Section \ref{sec-formulas}. It is indeed enough to consider the $V=1$ case because the Epstein zeta function is homogeneous, i.e.
\begin{equation}\label{homog}
\zeta_{V^{\frac{1}{3}}L}(s)=V^{-\frac{s}{3}}\zeta_L(s), \quad \forall V>0, \quad\forall s\in \R.
\end{equation}

It is already known (see \cite{Ennola,Gruber}) that $\mathsf{D}_3$ and $\mathsf{D}_3^*$ are both local minimizers of $L\mapsto \zeta_L(s)$ in $\mathcal{L}_3(1)$ for all $s>0$. Except of the minimality result of $\mathsf{D}_3$ for the height of the flat torus $L\mapsto \zeta_L'(0)$ in \cite{SarStromb} (and its very recent application \cite{RenWeiBCC} to the optimality of $\mathsf{D}_3^*$ for diblock copolymer molecular systems) as well as the optimality of $\Z^d$ among $d$-dimensional orthorhombic lattices following from Montgomery's work \cite{Mont}, we are not aware of any other global optimality results in dimension $d=3$ for the Epstein zeta function.

First, we have re-checked the following well-known observations (see \cite{SarStromb}):
\begin{itemize}
\item for all $0<s<\frac{3}{2}$, $\mathsf{D}_3^*$ is the unique minimizer of $L\mapsto \zeta_L(s)$ in $\mathcal{L}_3(1)$;
\item for all $s>\frac{3}{2}$, $\mathsf{D}_3$ is the unique minimizer of $L\mapsto \zeta_L(s)$ in $\mathcal{L}_3(1)$.
\end{itemize}
Furthermore, it is easy to show (see e.g. \eqref{ebfcctrans} applied to $s=3/2$) that $\zeta_{\mathsf{D}_3}\left(\frac{3}{2}\right)=\zeta_{\mathsf{D}_3^*}\left(\frac{3}{2}\right)$.

\begin{figure}[!h]
	\centering
	\includegraphics[clip,width=8cm]{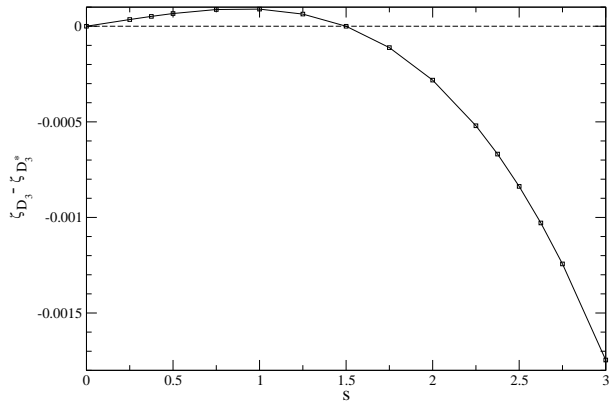}
	\caption{Plot of $s\mapsto \zeta_{\mathsf{D}_3}(s)-\zeta_{\mathsf{D}_3^*}(s)$ for $s\in [0,3]$.}
	\label{ef-bcc-s}
\end{figure}

\paragraph{\textbf{Minimization for $0<s<3$.}} Here we use the analytic continuation with respect to $s$ of the Epstein zeta function. The energies of $\mathsf{D}_3$ and $\mathsf{D}_3^*$ have close values, hence we plot their difference in Figure \ref{ef-bcc-s}.
It vanishes for $s\to 0^+$ as $\zeta_L(s)\to -1/2$ for any lattice $L$
and it remains finite for $s\to 3^-$ by the general Kronecker's Limit Formula \cite[Thm. 2.2]{Chiu}.
The latter value was calculated semi-analytically.
We can see that $\zeta_{\mathsf{D}_3}(s)-\zeta_{\mathsf{D}_3^*}(s)$ is positive for $0<s<3/2$ and negative otherwise, as expected.

\paragraph{\textbf{Minimization for $s>3$.}}

Recall that for large $s$ the FCC and HCP lattices could compete as they are closed packed lattices so that they should have common asymptotes.
The large $s$ expansions can be deduced from summation
formulas of the type \eqref{ehcp} by inspection. 
For the FCC lattice we get

\begin{equation}\label{efccbigs}
\zeta_{\mathsf{D}_3}(s)\approx\frac{1}{2^{s/6}}\left[6+\frac{3}{2^{s/2}} +\frac{12}{3^{s/2}}+\frac{6}{2^{s}}+\ldots\right]
\end{equation}
whereas for the HCP one

\begin{equation}\label{ehcpbigs}
\zeta_{\mathsf{A3}}(s)\approx\frac{1}{2^{s/6}}\left[6+\frac{3}{2^{s/2}} +\frac{1}{\big(\frac{8}{3}\big)^{s/2}}+\frac{9}{3^{s/2}}+\ldots\right].
\end{equation}

\begin{figure}[!h]
	\centering
	\includegraphics[clip,width=8cm]{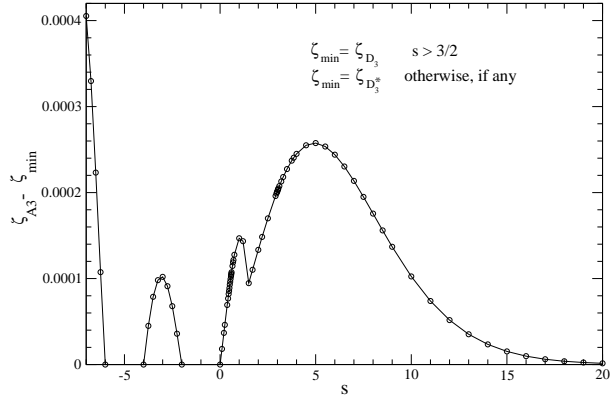}
	\caption{Plot of $s\mapsto \zeta_{\mathsf{A3}}(s)-\min\{\zeta_{\mathsf{D}_3}(s),\zeta_{\mathsf{D}_3^*}(s)\}$ for $s\in [-7,20]$.}
	\label{ehcp-emins}
\end{figure}

We can see that both the leading and sub-leading terms are equal and for large enough $s$ the third term decides that $\zeta_{\mathsf{D}_3}(s)<\zeta_{\mathsf{A3}}(s)$ as $3>8/3$.
This still does not mean that HCP could not prevail for some medium values of $s$. 
The values of the third term become equal for $s\approx 21$. 
Hence we performed high precision numerical calculations presented in Figure \ref{ehcp-emins}.

We can see that $\zeta_{\mathsf{A3}}(s)-\zeta_{\mathsf{D}_3}(s)$ remains positive for all $s>3/2$, as expected.
The apparent non-analyticity at $s=3/2$ is given by the fact, that we subtract another function for $s<3/2$, namely $\zeta_{\mathsf{A3}}(s)-\zeta_{\mathsf{D}_3^*}(s)$, in order to show that $\mathsf{A3}$ does not become the minimizer at any $s$.
Note that the plotted difference is smooth at $s\to 3$ although both $\zeta_{\mathsf{A3}}(s)$ and $\zeta_{\mathsf{D}_3}(s)$ diverge.

\paragraph{\textbf{Minimization for $s<0$.}}

\begin{figure}[!h]
	\centering
	\includegraphics[clip,width=8cm]{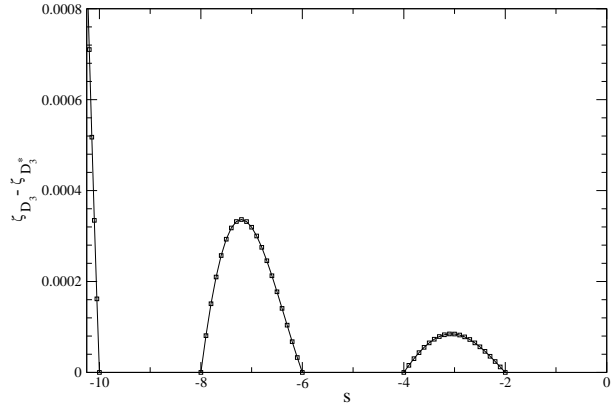}
	\caption{Plot of $s\mapsto\zeta_{\mathsf{D}_3}(s)-\zeta_{\mathsf{D}_3^*}(s)$ when $s\in [-11,0)$.}
	\label{efcc-ebcc_slt0}
\end{figure}

 Here we use the analytic continuation with respect to $s$ of the Epstein zeta function. At first let us state that this region is physically not very reasonable, but we can study it nevertheless. As recalled in the introduction, its minimum value is related to the Jellium problem with Riesz interaction.
All analytically continued energies include the factor $1/\Gamma(s/2)$.
It tends to zero at $s/2\to 0,-1,-2,$ etc.
The case $s\to 0$ should be omitted, as we already mentioned that, for all lattice $L$, $\zeta_L(s)\to -1/2$ in the sense of limit.
All the other zeros are present; in the 1D case of Riemann's zeta function they are called trivial zeros.
The energies are analytical functions thus they change signs at these points; they become negative at $-2<s<0$, $-6<s<-4$, etc., and they are positive otherwise.
The negativity of energies gives rise to extremely low possible values of energies, say for simple hexagonal lattice $\mathsf{A_f}$ with very large anisotropy parameter $\Delta=d/a$ (where $a$ is the lattice constant and $d$ is the inter-layer distance), or for very dilated orthorhombic lattices, and it has no lower bound.
In other words for $-4n-2<s<-4n$ with $n=0,1,2, \ldots$ there is no lattice the single minimizer as the energies diverge $\zeta_{\mathsf{A_f}}(s)\to -\infty$ for $\Delta\to \infty$.
Finally for intervals $-4n<s<-4n+2$ with $n=1,2, \ldots$ where the energies are positive we have the minimizer and it is the BCC lattice.
This was used in the left part of Figure \ref{ehcp-emins} as well as in Figure \ref{efcc-ebcc_slt0}.
One could plot those differences also in the intervals where they are negative and see continuous oscillating functions, but non of the lattices represents the minimizer there.

Concluding, for real $s\ne 3$ we still have two lattices that become minimizers at some intervals of $s$.
There is no minimizer for $-4n-2<s<-4n$ with $n=0,1,2, \ldots$; the FCC lattice minimizes the Riesz energy for $s>3/2$ and BCC otherwise. The result for negative $s$ is new.

According to our numerical investigation, we can therefore write the following conjecture for the Riesz energy, completing the one of Sarnak and Str\"ombergsson in \cite{SarStromb}.
\begin{Conjecture} We have the following minimizers in $\widetilde{\mathcal{L}_3}(1)$ for the Epstein zeta function $L\mapsto \zeta_L(s)$:
\begin{itemize}
\item for all $n\in \N$ and all $s\in (-4n,-4n+2)$, the unique minimizer is $\mathsf{D}_3^*$;
\item for all $n\in \N\cup\{0\}$ and all $s\in (-4n-2,-4n)$, the lattice energy has no minimizer;
\item for all $s\in (0,3/2)$, the unique minimizer is $\mathsf{D}_3^*$;
\item for all $s>3/2$, the unique minimizer is $\mathsf{D}_3$.
\end{itemize}
\end{Conjecture}

\section{Minimizers of the Lennard-Jones type energy}\label{sec-LJ}

 In this section, we numerically investigate the minimizers of the Lennard-Jones type energy $L\mapsto E_{n,m}^{\rm LJ}[L]$ in  $\mathcal{L}_3$ or $\widetilde{\mathcal{L}_3}$ (i.e. without any constraint on the density) and $\mathcal{L}_3(V)$ or $\widetilde{\mathcal{L}_3}(V)$, $V>0$ (i.e. at fixed density). Using the homogeneity \eqref{homog} of the Epstein zeta function, we recall that, for all $V>0$ and all $L\in \widetilde{\mathcal{L}_3}(1)$,
 $$
 E_{n,m}^{\rm{LJ}}[V^{\frac{1}{3}}L]=\frac{m}{n-m}\frac{\zeta_L(n)}{V^{\frac{n}{3}}}-\frac{n}{n-m}\frac{\zeta_L(m)}{V^{\frac{m}{3}}}.
 $$
For making our plots, we therefore use again the formulas derived in Section \ref{sec-formulas} evaluated by using the symbolic computer language {\it Mathematica} with precision of 16 decimal digits.

Again, only local minimality results for $E_{n,m}^{\rm LJ}$ are known. In particular, $V^{\frac{1}{3}}\mathsf{D}_3$ and $V^{\frac{1}{3}}\mathsf{D}_3^*$ have been shown in \cite{Beterminlocal3d} to be locally minimal for $E_{n,m}^{\rm LJ}$ in $\mathcal{L}_3(V)$ for all fixed $V\in (0,V_0)$ where $V_0=V_0(n,m)$ is explicit and sharp. Furthermore, the SC lattice $V^{\frac{1}{3}}\Z^3$ is also showed in the same work to be locally minimal in a certain interval $(V_1,V_2)$ of volume. The only other known results are the asymptotic global optimality of a FCC lattice for large $s$ (see \cite{OptinonCM}) and some other minimality results among lattices with prescribed number of nearest-neighbors in \cite{LBbonds21}.  

The values of the Lennard-Jones exponents $m$ and $n$,
constrained by $m<n$, can lie in three distinct regions: $(3,\infty)$,
$[0,3)$ and $(-\infty,0)$. While the lattice sums are convergent in
the region $(3,\infty)$, they diverge (but still individual terms go
asymptotically to zero) in the ``critical'' strip $[0,3)$ and
they diverge (as well as individual terms diverge asymptotically)
on the half-line $(-\infty,0)$.
In this paper, we consider only such cases when both Lennard-Jones exponents
$m$ and $n$ lie in the same region.

\subsection{$3<m<n$}

\begin{figure}[!h]
	\centering
\includegraphics[clip,width=8cm]{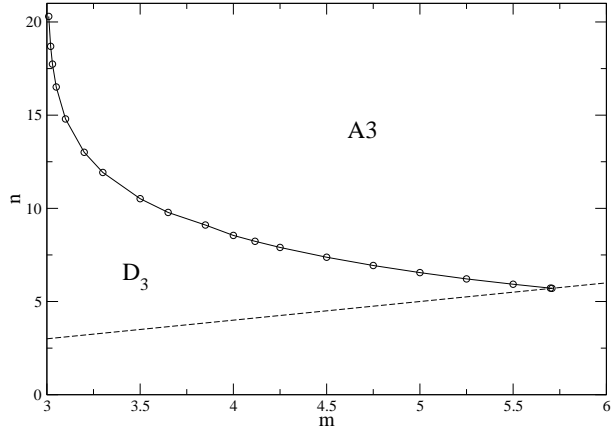}
\caption{Global minimizer o $E_{n,m}^{\rm{LJ}}$ in $\widetilde{\mathcal{L}_3}$ with respect to $(n,m)$. The dash line corresponds to $m=n$. The phase diagram for the minimization problem on  $\mathcal{L}_3$ is obtained by replacing the HCP phase by a FCC one.}
\label{phasediagn-m}
\end{figure}

\textbf{ Minimization in $\mathcal{L}_3$ and $\widetilde{\mathcal{L}_3}$.} Using dimension reduction techniques developed in \cite{OptinonCM} for Lennard-Jones type energies, we investigate the minimum of $E_{n,m}^{\rm LJ}$ in $\mathcal{L}_3$. As already explained in \cite{ModifMorse}, only two lattices appear to be global minimizers of this energy: the FCC lattice and the HCP structure (see Figure \ref{phasediagn-m}). It seems that a large (resp. narrow) well, i.e. when $n-m$ is small (resp. large) favors the FCC (resp. HCP) lattice as a ground state. It has to be noticed that, whatever $n$ and $m$ are, the global minimizer of $E_{n,m}^{\rm LJ}$ in $\mathcal{L}_3$ is a FCC lattice. This also supports the conjecture that we stated in \cite{Beterminlocal3d}.

\begin{figure}[!h]
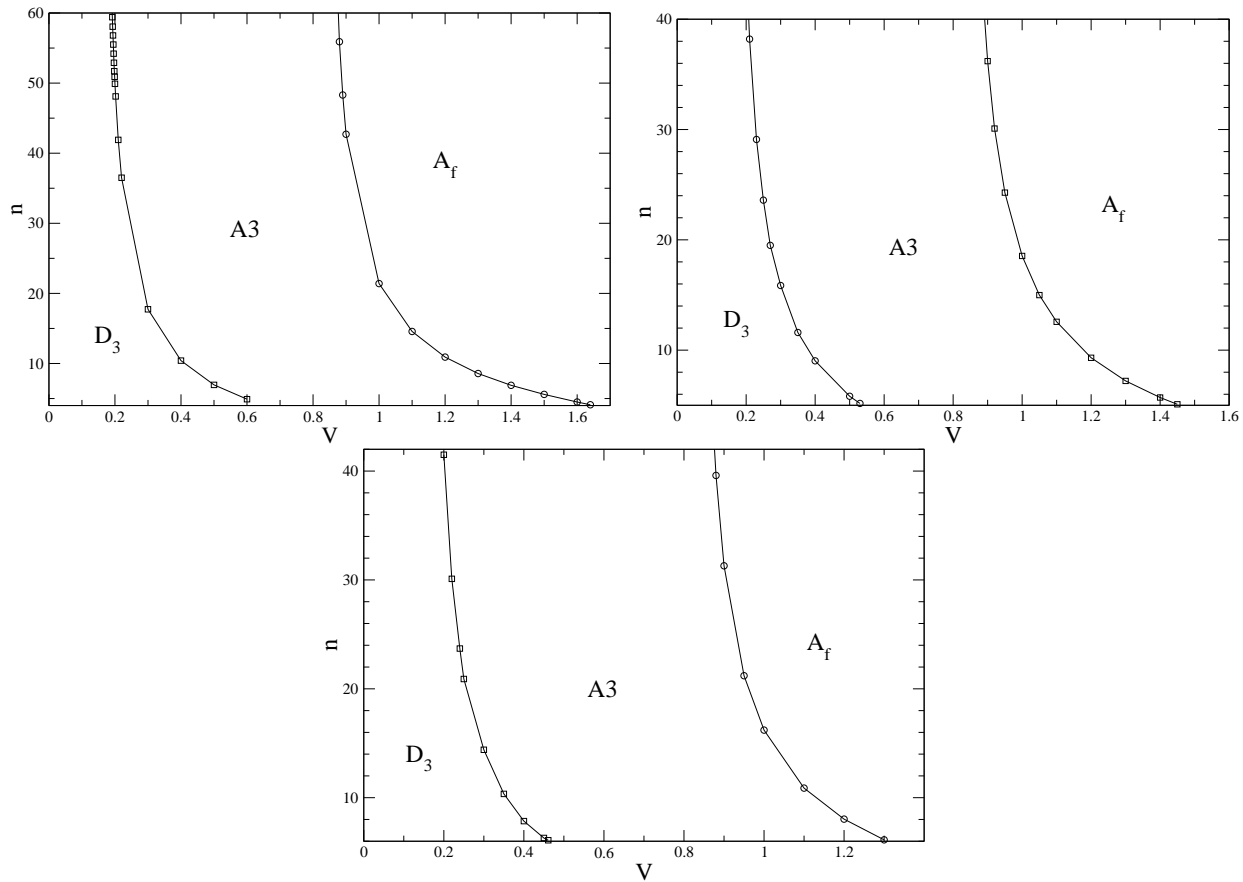

	\centering
\includegraphics[clip,width=8cm]{phasediagn-4V}\quad \includegraphics[clip,width=8cm]{phasediagn-5V2}\\ \includegraphics[clip,width=8cm]{phasediagn-6V2}
\caption{Phase diagram of the minimizer of $E_{n,m}^{\rm{LJ}}$ in $\widetilde{\mathcal{L}_3}(V)$ for $m=4$ (up left) and $m=5$ (up right) and $m=6$ (down), $n$ varying. The phase diagram for the minimization problem on  $\mathcal{L}_3(V)$ is roughly obtained by replacing the HCP phase by a FCC one.}
\label{phdiagn-4v}
\end{figure}

\begin{figure}[!h]
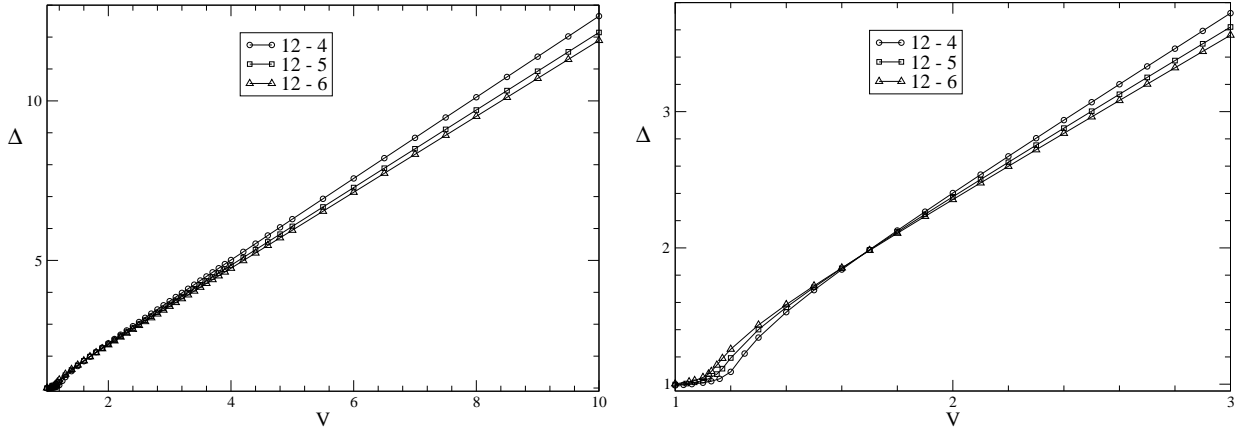

	\centering
\includegraphics[clip,width=8cm]{delta-V}\quad \includegraphics[clip,width=8cm]{delta-Vdetail}
\caption{Simple Hexagonal phase. Plot of $V\mapsto \argmin_\Delta E_{n,m}^{\rm{LJ}}[V^{\frac{1}{3}}\mathsf{A}_f(\Delta)]$ for $(n,m)\in \{(12,m) : m=4,5,6 \}$.}
\label{delta-v}
\end{figure}

\textbf{Minimization in $\mathcal{L}_3(V)$ and $\widetilde{\mathcal{L}_3}(V)$.} For the problem of minimizing $E_{n,m}^{\rm LJ}$ at fixed density, we have plotted in Figure \ref{phdiagn-4v} the phase diagram of $(n,V)\mapsto \argmin_{L\in \mathcal{L}_3(V)} E_{n,m}^{\rm LJ}[L]$ where $m\in \{4,5,6\}$. We observe that there are only three kinds of minimizers in $\widetilde{\mathcal{L}_3}(V)$: the FCC lattice $V^{\frac{1}{3}}\mathsf{D}_3$, the Simple Hexagonal lattice $V^{\frac{1}{3}}\mathsf{A_f}(\Delta(V))$ and the HCP structure $V^{\frac{1}{3}}\mathsf{A3}$. More precisely, there exist $V_{n,m},V_{n,m}'>0$ such that:
\begin{itemize}
\item for all $V\in (0, V_{n,m})$, $V^{\frac{1}{3}}\mathsf{D}_3$ is the unique minimizer of $E_{n,m}^{\rm LJ}$ in $\widetilde{\mathcal{L}_3}(V)$;
\item for all $V\in (V_{n,m}, V_{n,m}')$, $V^{\frac{1}{3}}\mathsf{A3}$ is the unique minimizer of $E_{n,m}^{\rm LJ}$ in $\widetilde{\mathcal{L}_3}(V)$;
\item for all $V> V_{n,m}'$, $V^{\frac{1}{3}}\mathsf{A_f}(\Delta)$ is the unique minimizer of $E_{n,m}^{\rm LJ}$ in $\widetilde{\mathcal{L}_3}(V)$ for some $\Delta=\Delta(V)$, where the behavior of $\Delta(V)$ is depicted in Figure \ref{delta-v} for $n=12$ and $m\in \{4,5,6\}$.
\end{itemize}
If we minimize $E_{n,m}^{\rm LJ}$ only in $\mathcal{L}_3(V)$, the HCP phase is replaced by a FCC one in such a way that 
\begin{itemize}
\item for all $V\in (0,V_{n,m}')$, $V^{\frac{1}{3}}\mathsf{D}_3$ is the unique minimizer of $E_{n,m}^{\rm LJ}$ in $\mathcal{L}_3(V)$;
\item for all $V> V_{n,m}'$, $V^{\frac{1}{3}}\mathsf{A_f}(\Delta)$ is the unique minimizer of $E_{n,m}^{\rm LJ}$ in $\mathcal{L}_3(V)$ for some $\Delta=\Delta(V)$ as depicted in Figure \ref{delta-v}.
\end{itemize}



In \cite{Beterminlocal3d}, the interval of volumes $V$ where the SC lattice is locally minimal for $E_{n,m}^{\rm{LJ}}$ in $\mathcal{L}_3(V)$ has been derived. Here we have observed that a family of SH lattices have lower energy than the SC one. This is new compared to the two-dimensional case where the square lattice, that has been shown in \cite{Beterloc} to be locally minimal in some interval of volume, is also observed to be minimal in $\mathcal{L}_2(V)$ in the same interval. Furthermore, our numerical findings concerning the global minimizer of $E_{n,m}^{\rm{LJ}}$ in $\mathcal{L}_3$ support \cite{Beterminlocal3d} and confirm \cite{ModifMorse}.

\subsection{\textbf{$0<m<n<3$}}
\begin{figure}[t]
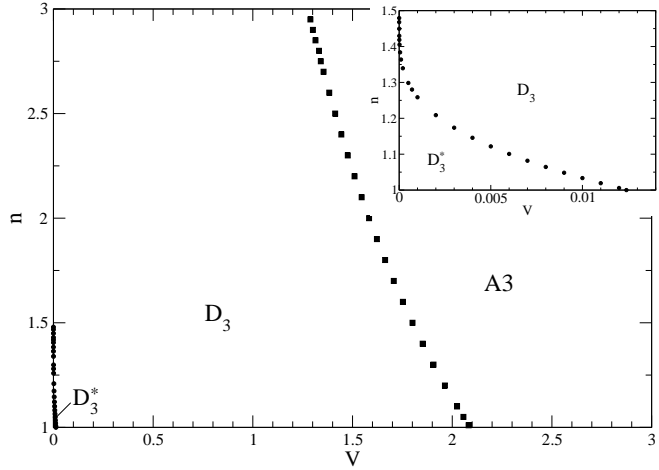

\setbox1=\hbox{\includegraphics[clip,height=6.2cm]{phdiagn-1.eps}}
\includegraphics[clip,height=6.2cm]{phdiagn-1.eps}\llap{\raisebox{3.35cm}{\includegraphics[clip,height=2.8cm]{phdn-1det.eps}}}
\caption{Phase diagram for the minimum Lennard-Jones energy for the exponents
$m=1$ and $1<n<3$ at fixed density $1/V$.}
\label{ep1b}
\end{figure}

Let us now consider that the Lennard-Jones exponents $n$ and $m$ lie in
the critical strip and fix $m=1$.
It turns out that for any value of $1<n<3$ the ${\rm A_f}$
lattice with the extreme value of the anisotropy parameter
$\Delta_{\rm opt}\to\infty$ yields the lowest possible energy per site
$E_{\rm A_f}\to -\infty$. 
In order to get non-trivial results at fixed density, we will restrict
ourselves to rigid lattices: ${\Z^3}$, ${\rm D_3^*}$, ${\rm D_3}$
and ${\rm A3}$.
One can see in the inset of Fig.~\ref{ep1b} that for small values of $n$
close to 1 and very small elementary cell volume $V$ up to 0.01,
the ${\rm D_3^*}$ lattice prevails.
Note that the phase boundary between ${\rm D_3^*}$ and ${\rm D_3}$ lattices
ends at $n=1.5$; this is due to the fact that for an infinitesimal $V$
the $n$ term dominates in the Lennard-Jones energy and it holds that
$E_{\rm D_3^*}(3/2)=E_{\rm D_3}(3/2)$ for the Riesz interaction at $s=3/2$.
For larger values of $n$ and $V$, there is a competition between
the ${\rm D_3}$ and ${\rm A3}$ lattices as minimizers of the energy.
The boundary between the two lattices runs within the interval approximately $1.28 < V < 2.09$.
There is numerical evidence that if we chose the exponent $m\ge 1.5$,
the ${\rm D_3^*}$ lattice would disappear from the phase diagram.

\subsection{\textbf{$-2<m<n<0$}}
Within this interval of the exponents, the ${\rm D_3^*}$ lattice becomes
the minimizer of the enegy for any value of $V$. 
The ${\rm A_f}$ lattice has its (finite) minimum energy for a finite value
of $\Delta_{\rm opt}$,
but its energy is not low enough to become the global minimizer.

\subsection{\textbf{$-4<m<n<-2$}}
\begin{figure}[t]
\includegraphics[width=8cm]{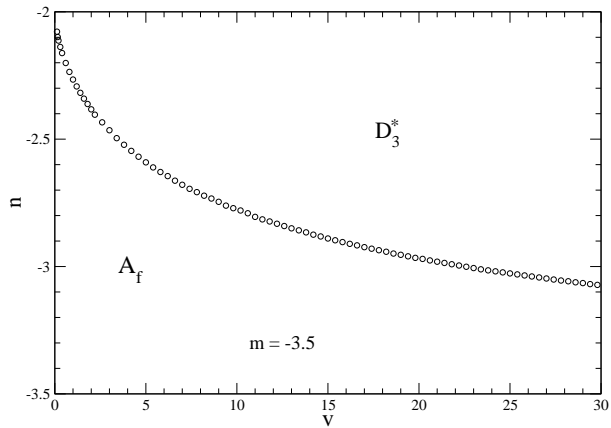}
\caption{Phase diagram of the lowest Lennard-Jones energy for the exponents
$m=-3.5$ and $-3.5<n<-2$.}
\label{bcchex}
\end{figure}

Now both ${\rm D_3^*}$ and ${\rm A_f}$ lattices are candidates for global
minimizers, the latter lattice with a finite parameter $0.6<\Delta_{opt}<0.9$
for the presently calculated data.
The typical phase diagram for the Lennard-Jones exponents $m=-3.5$ and
$-3.5<n<-2$ is plotted in Fig.~\ref{bcchex}.
The phase curve between the ${\rm D_3^*}$ and ${\rm A_f}$ lattices
approaches to $n=-3.5$ at asymptotically large $V\to\infty$.
Analogous phase diagrams appear in intervals $-6<m<n<-4$, etc.

According to our numerical study, we can write the following conjecture for the ground state of the Lennard-Jones type energy.
\begin{Conjecture}
Concerning the global minimizer of $E_{n,m}^{\textnormal{LJ}}$, we have:
\begin{itemize}
\item for $3<m<n$, then we have two cases:
\begin{enumerate}
\item in $\mathcal{L}_3$, the unique minimizer of $E_{n,m}^{\textnormal{LJ}}$ is $\mathsf{D}_3$, up to rescaling.
\item in $\widetilde{\mathcal{L}_3}$, there exists $m_0\approx 5.7$ such that, 
\begin{itemize}
\item if $m<m_0$, then there exists $n_m$ such that if $n<n_m$ (resp. $n>n_m$), the unique minimizer of $E_{n,m}^{\textnormal{LJ}}$ is $\mathsf{D}_3$ (resp. $\mathsf{A}_3$), up to rescaling.
\item if $m>m_0$, then the unique minimizer of $E_{n,m}^{\textnormal{LJ}}$ is $\mathsf{A}_3$.
\end{itemize}
\end{enumerate}
\item for $0<m<n<3$, $E_{n,m}^{\textnormal{LJ}}$ does not have any minimizer at fixed density.
\item for all $k\in \N\cup \{0\}$, for all $-4k-2<m<n<-4k$, the unique minimizer of $E_{n,m}^{\textnormal{LJ}}$ in $\widetilde{\mathcal{L}_3}$ is $\mathsf{D}_3^*$, up to rescaling;
\item for all $k\in \N$, for all $-4k<m<n<-4k+2$, $\mathsf{D}_3^*$ and $\mathsf{A_f}$ are both minimizers, up to rescaling, of $E_{n,m}^{\textnormal{LJ}}$ in $\widetilde{\mathcal{L}_3}$.
\end{itemize}
\end{Conjecture}

\section{Explicit formulas for three-dimensional Epstein zeta functions} \label{sec-formulas}
This section is devoted to the construction of explicit formulas for
Epstein zeta functions associated to the three-dimensional lattices
considered in the present paper.
We start with converging sums, i.e., the parameter $s$ is larger than dimension
3, and express them as integrals over powers of Jacobi elliptic functions.
Then using specific tricks with elliptic functions these integrals are
rewritten into the ones which represent an analytic continuation of
Epstein zeta functions to the whole complex plane except for $s= 3$.

\subsection{Simple cubic lattice}
Let particles be localized at sites of the SC lattice with spacing $a$.
In this paper, we compare the energy per particle for various 3D lattice
structures at a fixed density of particles, say the unit one.
The particle density associated with the SC lattice, given by $\rho=1/a^3$,
is equal to one when $a=1$.
Particles interact pairwisely by the Riesz potential $1/r^s$ where
$r=\vert {\bf r}\vert$ is the distance between two particles and $s$
is a complex number.
The energy per particle is given by the Epstein zeta function associated
with the SC lattice
\begin{equation} \label{zetaSC}
\zeta_{\mathbb{Z}^3}(s) = \frac{1}{2}\sum_{(j,k,l) \ne (0,0,0)}
\frac{1}{(j^2+k^2+l^2)^{s/2}} , 
\end{equation}
where the prefactor $1/2$ is due to the fact that every energy is shared
by two particles and the sum converges only if the real part of $s$
$\Re(s)>3$.
Using the $\Gamma$-identity
\begin{equation} \label{gamma}
\frac{1}{r^s} = \frac{1}{(r^2)^{s/2}}
=\frac{1}{\Gamma(s/2)} \int_0^\infty {\rm d} t\, t^{\frac{s}{2}-1}
{\rm e}^{-r^2 t} ,
\end{equation}
the expression (\ref{zetaSC}) can be rewritten as
\begin{eqnarray}
\zeta_{\mathbb{Z}^3}(s) & = & \frac{1}{2\Gamma(s/2)}
\int_0^\infty \frac{{\rm d} t}{t} t^{s/2}  
\left[\sum_{j,k,l=-\infty}^{\infty} {\rm e}^{-(j^2+k^2+l^2)t}-1\right] \nonumber \\
& = & \frac{1}{2\Gamma(s/2)} \int_0^\infty \frac{{\rm d} t}{t}
t^{s/2}  \left[ \theta_3^3\left({\rm e}^{-t}\right)-1 \right] 
\nonumber \\
& = & \frac{\pi^{s/2}}{2\Gamma(s/2)} \int_0^\infty \frac{{\rm d} t}{t}
t^{s/2}  \left[ \theta_3^3\left({\rm e}^{-\pi t}\right)-1 \right] ,
\label{esc}
\end{eqnarray}
where the Jacobi elliptic function with zero argument
$\vartheta_3(0,q) \equiv \theta_3(q) = \sum_{j=-\infty}^{\infty} q^{j^2}$
\cite{Gradshteyn} was introduced for $q=e^{i\pi \tau}$, $\Im(\tau)>0$. Notice that \eqref{esc} can be obtained from the Mellin transform of the Jacobi theta function (see e.g. \cite{JL} and \cite[Sec. 1.4.2]{Terras1}).
With regard to the definition of $\theta_3(q)$, it holds that
$\theta_3\left( {\rm e}^{-\pi t} \right) \sim_{t\to\infty}
1 + 2 {\rm e}^{-\pi t} + \cdots$
and the integral in (\ref{esc}) converges at large $t$ for any $s$.
Using the Poisson summation formula
\begin{equation} \label{PSF}
\sum_{j=-\infty}^{\infty} {\rm e}^{-(j+\phi)^2\pi t} = \frac{1}{\sqrt{t}}
\sum_{j=-\infty}^{\infty} {\rm e}^{2\pi{\rm i}j\phi} {\rm e}^{-(\pi j^2)/t} 
\end{equation}
it can be shown that
\begin{equation} \label{sing3}
\theta_3\left( {\rm e}^{-\pi t} \right) \mathop{\sim}_{t\to 0} \frac{1}{\sqrt{t}}
\left( 1 + 2 {\rm e}^{-\pi/t} + \cdots \right) .
\end{equation}
The function under integration in (\ref{esc}) behaves like $t^{\frac{s-3}{2}-1}$
for $t\to 0$, so the real part of $s$ must be greater than 3 as was expected.

To derive an analytic continuation of (\ref{esc}) to the complex-$s$ plane, 
we follow the derivation presented in Ref. \cite{Zeros} for $d$-dimensional
hypercubic lattices.
In the integral on the rhs of (\ref{esc}), we split the integration over $t$
into intervals $[0,1]$ and $[1,\infty]$.
In the second integral over $t\in [1,\infty]$, one applies the equality
\begin{equation}
\theta_3(e^{-\pi t}) = \frac{1}{\sqrt{t}} \theta_3\left({\rm e}^{-\pi/t}\right)
\end{equation}  
which follows from (\ref{PSF}) and substitutes $t=1/t'$ to get
\begin{equation} \label{upperint}
\int_1^\infty \frac{{\rm d}t}{t} t^{s/2}
\left[ \theta_3^3\left( {\rm e}^{-\pi t}\right)-1 \right] =
\int_0^1 \frac{{\rm d}t'}{t'} {t'}^{(3-s)/2}
\left[\theta_3^3\left( {\rm e}^{-\pi t'}\right)-{t'}^{-3/2} \right].
\end{equation}
As concerns the first integral over $t\in [0,1]$, to subtract the singularity
of $\theta_3^3\left({\rm e}^{-\pi t}\right) \sim t^{-3/2}$ as $t\to 0$ one adds
the vanishing contribution $-t^{-3/2}+t^{-3/2}$ and integrates out the remaining
terms proportional to $t^{-3/2}-1$, with the result
\begin{equation} \label{escuni}
\frac{\Gamma{(s/2)}}{\pi^{s/2}} \zeta_{\mathbb{Z}^3}(s)
= -\frac{1}{s} - \frac{1}{3-s} +\int_{0}^1 \frac{{\rm d}t}{2t}\
\left[ t^{s/2}+t^{(3-s)/2}\right]
\left[\theta_3^3\left( {\rm e}^{-\pi t}\right)-\frac{1}{t^{3/2}}\right] .
\end{equation}
The difference in the square bracket is proportional to
$t^{-3/2} e^{-\pi/t}$ for $t\to 0$ and hence the formula (\ref{escuni}) converges
for all complex $s$, except for the singular point $s=3$.
The limit $s\to 0$ is not problematic since the singularity $-1/s$ on
the rhs of (\ref{escuni}) has a counterpart $\Gamma(s/2)\sim 2/s$
on the lhs, so that $\zeta_{\mathbb{Z}^3}(0)=-1/2$.
The numerical evaluation of $\zeta_{\mathbb{Z}^3}(s)$ by using (\ref{escuni})
at one point $s$ takes around 5 seconds of CPU time on a standard PC,
the achieved accuracy is around 29 decimal digits.
Similar characteristics occur in the numerical evaluation of all
Epstein functions studied below, except for the HCP structure given by the lattice sum for which the calculation of the Epstein function takes around 1 minute.

It is known that the transformation $s\to 3-s$ connects the lattice sums of
dual (reciprocal) lattice structures \cite{Latticesums}. 
The SC lattice is self-dual.
The rhs of (\ref{escuni}) is invariant with respect to this transformation,
hence it holds that
\begin{equation} \label{esctrans}
\zeta_{\mathbb{Z}^3}(s) = \frac{\pi^{s-3/2}\,
\Gamma\left(\frac{3-s}{2}\right)}{\Gamma\left(\frac{s}{2}\right)}
\zeta_{\mathbb{Z}^3}(3-s) .
\end{equation}

\subsection{Body centered cubic lattice}
The BCC lattice is composed of two SC lattices, shifted with respect to one
another by half-period along each of the three coordinates.
There are two particles per elementary cube of side $a_{\rm BCC}$,
i.e., $\rho=2/a^3_{\rm BCC}=1$ which implies that the unit density corresponds
to $a_{\rm BCC}=2^{1/3}$.
Consequently, 
\begin{equation} \label{ebcc1}
\zeta_{\mathsf{D}_3^*}(s) = \frac{1}{2^{s/3}}
\left[ \zeta_{\mathbb{Z}^3}(s) + \psi(1/2,1/2,1/2;s) \right],
\end{equation}
where
\begin{eqnarray} 
\psi(x_1,x_2,x_3;s) & = & \frac{1}{2} \sum_{j,k,l=-\infty}^{\infty}
\frac{1}{[(j-x_1)^2+(k-x_2)^2+(l-x_3)^2]^{s/2}} \nonumber \\
& = & \frac{1}{2\Gamma(s/2)} \int_0^\infty \frac{{\rm d} t}{t} t^{s/2}
\sum_{j,k,l=-\infty}^{\infty} {\rm e}^{-[(j-x_1)^2+(k-x_2)^2+(l-x_3)^2]t} \label{psi}
\end{eqnarray}
is a sum over a lattice shifted with respect to the original cubic one.
Introducing another Jacobi elliptic function
$\theta_2(q)=\sum_{j=-\infty}^{\infty} q^{(j-\frac{1}{2})^2}$, defined for $q=e^{i\pi \tau}$, $\Im(\tau)>0$,
the BCC energy is expressible as
\begin{equation} \label{ebcc}
\zeta_{\mathsf{D}_3^*}(s)
= \frac{1}{2^{s/3}} \left[ \zeta_{\mathbb{Z}^3}(s)
+\frac{\pi^{s/2}}{2\Gamma(s/2)} \int_0^{\infty} \frac{{\rm d} t}{t} t^{s/2}
\theta_2^3({\rm e}^{-\pi t})\right] .
\end{equation}
In view of the Poisson summation formula (\ref{PSF}), the Jacobi theta
function $\theta_4(q)=\sum_{j=-\infty}^{\infty} (-1)^j q^{j^2}$ is ``dual'' to
$\theta_2$ in the sense that
\begin{equation} \label{theta24}
\theta_2\left( {\rm e}^{-\pi t}\right) = \frac{1}{\sqrt{t}}
\theta_4\left( {\rm e}^{-\pi/t}\right) .
\end{equation}
Thus,
\begin{equation}
\theta_2\left( {\rm e}^{-\pi t} \right) \mathop{\sim}_{t\to 0} \frac{1}{\sqrt{t}}
\left( 1 - 2 {\rm e}^{-\pi/t} + \cdots \right) 
\end{equation}
and the integral in (\ref{ebcc}) converges at small $t$ if $\Re(s)>3$
as was expected.

To derive an analytic continuation of (\ref{ebcc}) we proceed in
close analogy with the previous SC lattice.
The integration on the rhs of (\ref{ebcc}) is split into intervals
$[0,1]$ and $[1,\infty]$.
The second integral over $t\in [1,\infty]$ can be transformed by using
the relation (\ref{theta24}) and the consequent substitution $t=1/t'$ to
\begin{equation}\label{upperintbcc}
\int_1^\infty \frac{{\rm d}t}{t} t^{s/2} \theta_2^3({\rm e}^{-\pi t})
=\int_0^1 {\rm d}t'\, {t'}^{(1-s)/2} \theta_4^3({\rm e}^{-\pi t'}).
\end{equation}
The first integral over $t\in [0,1]$ is modified by adding the vanishing
contribution $-t^{-3/2}+t^{-3/2}$; the term $-t^{-3/2}$ removes the singularity
of $\theta_2^3({\rm e}^{-\pi t})$ as $t\to 0$ and the term $t^{-3/2}$
is integrated out.
The final formula for the Epstein function associated with the BCC lattice
reads as
\begin{eqnarray}
\frac{2^{s/3}\Gamma(s/2)}{\pi^{s/2}} \zeta_{\mathsf{D}_3^*}(s) & = &
\frac{\Gamma{(s/2)}}{\pi^{s/2}} \zeta_{\mathbb{Z}^3}(s) - \frac{1}{3-s} \nonumber \\
& & + \int_0^1 \frac{{\rm d} t}{2 t} \left\{ t^{s/2} \left[
\theta_2^3\left( {\rm e}^{-\pi t}\right) - \frac{1}{t^{3/2}}\right] +
t^{(3-s)/2} \theta_4^3\left( {\rm e}^{-\pi t}\right) \right\} ,
\label{ebccuni}
\end{eqnarray}
where $\zeta_{\mathbb{Z}^3}(s)$ is given by (\ref{escuni}).
Since
\begin{equation}
\theta_4\left( {\rm e}^{-\pi t}\right) \mathop{\sim}_{t\to 0}
\frac{1}{\sqrt{t}} {\rm e}^{-\pi/(4 t)} + \cdots ,  
\end{equation}
the integral in (\ref{ebccuni}) converges for any complex $s$.

\subsection{Face centered cubic lattice}
There are four particles per elementary cube of side $a_{\rm FCC}$.
thus $\rho=4/a^3_{\rm FCC}=1$ which implies that $a_{\rm FCC}=2^{2/3}$.
For $\Re(s)>3$, one gets
\begin{eqnarray}
\zeta_{\mathsf{D}_3}(s) & = & \frac{1}{2^{2s/3}}
\left[ \zeta_{\mathbb{Z}^3}(s) + \psi\left(0,\frac{1}{2},\frac{1}{2};s\right)
+\psi\left(\frac{1}{2},0,\frac{1}{2};s\right)
+\psi\left(\frac{1}{2},\frac{1}{2},0;s\right) \right] \nonumber \\ 
& = & \frac{1}{2^{2s/3}} \left[ \zeta_{\mathbb{Z}^3}(s) +
\frac{3\pi^{s/2}}{2\Gamma(\frac{s}{2})}
\int_0^\infty \frac{{\rm d} t}{t} t^{s/2} \theta_2^2\left( {\rm e}^{-\pi t}\right)
\theta_3\left( {\rm e}^{-\pi t}\right) \right] . \label{efcc}
\end{eqnarray}
Splitting the integration into the intervals $[0,1]$ and $[1,\infty]$
and using the equality
\begin{equation}\label{upperintfcc}
\int_1^{\infty} \frac{{\rm d}t}{t} t^{s/2}
\theta_2^2\left( {\rm e}^{-\pi t}\right) \theta_3\left( {\rm e}^{-\pi t}\right)
= \int_0^1 {\rm d}t\, t^{(1-s)/2} \theta_4^2\left( {\rm e}^{-\pi t}\right)
\theta_3\left( {\rm e}^{-\pi t}\right) ,
\end{equation}
the analytic continuation of (\ref{efcc}) to the whole complex $s$-plane
(except for the singular point $s=3$) takes the form
\begin{eqnarray}
\frac{2^{2s/3} \Gamma(s/2)}{\pi^{s/2}} \zeta_{\mathsf{D}_3}(s) & = & 
\frac{\Gamma{(s/2)}}{\pi^{s/2}} \zeta_{\mathbb{Z}^3}(s) - \frac{3}{3-s}
\nonumber \\ & & + \frac{3}{2} \int_0^1 \frac{{\rm d} t}{t} \Bigg\{
t^{s/2} \left[ \theta_2^2\left( {\rm e}^{-\pi t}\right)
\theta_3\left( {\rm e}^{-\pi t}\right) - \frac{1}{t^{3/2}} \right]
\nonumber \\ & & + t^{(3-s)/2} \theta_4^2\left( {\rm e}^{-\pi t}\right)
\theta_3\left( {\rm e}^{-\pi t}\right) \Bigg\} . \label{efccuni}
\end{eqnarray}

The FCC and BCC lattices is a pair of dual structures.
Exploring theta-function identities \cite{WhiWat69}
\begin{equation} \label{ww}
\theta_3(q) = \theta_3(q^4)+\theta_2(q^4), \qquad
\theta_4(q) = \theta_3(q^4)-\theta_2(q^4) 
\end{equation}
in the representations (\ref{ebccuni}) and (\ref{efccuni})
it can be shown after some algebra that
\begin{equation}\label{ebfcctrans}
\zeta_{\mathsf{D}_3^*}(s) = \frac{\pi^{s-3/2}\,
\Gamma\left(\frac{3-s}{2}\right)}{\Gamma\left(\frac{s}{2}\right)}
\zeta_{\mathsf{D}_3}(3-s)
\end{equation}
or, vice versa, 
\begin{equation} \label{efbcctrans}
\zeta_{\mathsf{D}_3}(s) = \frac{\pi^{s-3/2}\,
\Gamma\left(\frac{3-s}{2}\right)}{\Gamma\left(\frac{s}{2}\right)}
\zeta_{\mathsf{D}_3^*}(3-s) .
\end{equation}
The turning point of these transformations is $s=3/2$ where
$\zeta_{\mathsf{D}_3}(3/2) = \zeta_{\mathsf{D}_3^*}(3/2)$.

\subsection{Simple hexagonal lattice}
The SH lattice is composed of parallel planes composed of the 2D hexagonal
lattice with spacing $a_{\rm SH}$, the distance between the nearest-neighbor
planes is $d$; the dimensionless parameter $\Delta=d/a_{\rm SH}$. 
The 2D hexagonal lattice can be considered as the union of two rectangle
lattices of sides $a_{\rm SH}$ and $\sqrt{3} a_{\rm SH}$, shifted with respect
to one another by a half-period.
Thus the sites of the SH lattice can be parameterized with respect to the
reference point $(0,0,0)$ as follows: $a_{\rm SH}(j,\sqrt{3}k,\Delta l)$
with integers $j,k,l$ such that $(j,k,l)\ne (0,0,0)$ and
$a_{\rm SH}\left[ (j+\frac{1}{2}),\sqrt{3}(k+\frac{1}{2}),\Delta l\right]$
with integer $j,k,l$.
There are two particles in the elementary rectangular parallelepiped
of volume $\sqrt{3} a_{\rm SH}^2 d$, so the unit density of particles
corresponds to $2/\left( \sqrt{3} \Delta a_{\rm SH}^3\right) =1$.
For $\Re(s)>3$, the energy per particle for the SH lattice is given by
\begin{eqnarray}
\zeta_{\mathsf{A_f}(\Delta)}(s) & = & \frac{\pi^{s/2}}{2 a_{\rm SH}^s \Gamma(s/2)}
\int_0^\infty \frac{{\rm d} t}{t} t^{s/2}
\Big[ \theta_3\left( {\rm e}^{-\pi t}\right)
\theta_3\left( {\rm e}^{-3\pi t}\right)
\theta_3\left( {\rm e}^{-\pi t\Delta^2}\right) \nonumber\\ & &
-1 + \theta_2\left( {\rm e}^{-\pi t}\right)
\theta_2\left( {\rm e}^{-3\pi t}\right)
\theta_3\left( {\rm e}^{-\pi t\Delta^2}\right) \Big] .
\end{eqnarray}
The analytic continuation to the whole complex $s$-plane (except for
the point $s=3$) is obtained in the form
\begin{eqnarray}
\left( \frac{2}{\sqrt{3}\Delta} \right)^{s/3} \frac{\Gamma(s/2)}{\pi^{s/2}}
\zeta_{\mathsf{A_f}(\Delta)}(s) & = & - \frac{1}{s}
-\frac{2}{\Delta\sqrt{3}(3-s)} \nonumber \\ 
& & + \int_0^1 \frac{{\rm d} t}{2t}\Bigg\{ t^{s/2}
\Big[ \theta_3\left( {\rm e}^{-\pi t}\right)
\theta_3\left( {\rm e}^{-3\pi t}\right)
\theta_3\left( {\rm e}^{-\pi t\Delta^2}\right) \nonumber \\ 
& & + \theta_2\left( {\rm e}^{-\pi t}\right)
\theta_2\left( {\rm e}^{-3\pi t}\right)
\theta_3\left( {\rm e}^{-\pi t\Delta^2}\right)
- \frac{2}{\sqrt{3}\Delta t^{3/2}} \Big]\nonumber\\ 
& & + \frac{t^{(3-s)/2}}{\sqrt{3}\Delta}
\Big[ \theta_3\left( {\rm e}^{-\pi t}\right)
\theta_3\left( {\rm e}^{-\pi t/3}\right)
\theta_3\left( {\rm e}^{-\pi t/\Delta^2}\right) \nonumber \\ 
& & -\frac{\sqrt{3}\Delta}{t^{3/2}} + \theta_4\left( {\rm e}^{-\pi t}\right)
\theta_4\left( {\rm e}^{-\pi t/3}\right)
\theta_3\left( {\rm e}^{-\pi t/\Delta^2}\right)\Big]\Bigg\} . \nonumber\\
\end{eqnarray}

\subsection{Hexagonal close-packed lattice}
For very large $s\to \infty$ the Riesz potential $1/r^s$ tends to infinity
for $r<1$ and vanishes for $r>1$, i.e. it approaches the hard-sphere limit.
For this case a close-packed structure becomes the minimizer.
There are two such lattices in 3D, besides the FCC also the hexagonal
close-packed (HCP) lattice with the same packing ratio.
The two lattices have common first order asymptotic value of the Epstein zeta function -- since they have the same non-zero shortest vectors --
and it is relatively simple to derive how their difference behaves
for large $s$, see the relations \eqref{efccbigs} and \eqref{ehcpbigs}.

The HCP lattice is pictured in Fig. 1; the middle hexagonal layer
is shifted by a half-period with respect to the top and bottom
hexagonal lattices.
Denoting by $a_{\rm HCP}$ the spacing of the one-layer hexagonal lattice,
the distance between top and bottom hexagonal layers is given by
$d=\Delta a_{\rm HCP}$ with $\Delta = \sqrt{8/3}$.
As an elementary cell we take the rectangular parallelepiped
with sides $a_{\rm HCP}$ and $\sqrt{3} a_{\rm HCP}$ on the top and bottom
layers and the height $d$, its volume is given by
$\sqrt{3} a_{\rm HCP}^2 d = \sqrt{8} a_{\rm HCP}^3$.
There are 8 particles on the top and bottom layers each shared by 8 elementary
cells, 2 particles on the top and bottom layers each shared by 2 cells,
1 particle on the middle layer sitting in the cell interior and
2 particles on the middle layer sitting on the cell surface
(each shared by 2 cells), i.e. altogether there are 4 particles per
elementary cell. 
The density of particles $\rho=4/(\sqrt{8} a_{\rm HCP}^3)$ equals to
unity when $a_{\rm HCP} = 2^{1/6}$.

Particle positions on the top and bottom layers can be parameterized
with respect to the reference particle at the origin $(0,0,0)$
as follows: $a_{\rm HCP}(j,\sqrt{3}k,\Delta l)$ with integer $j,k,l$
such that $(j,k,l)\ne (0,0,0)$ and
$a_{\rm HCP}\left[ (j+\frac{1}{2}),\sqrt{3}(k+\frac{1}{2}),\Delta l\right]$
with integer $j,k,l$.
It can be shown that particle positions on the middle layer can be parameterized
as follows:
$a_{\rm HCP}\left[ j,\sqrt{3}(k+\frac{2}{3}),\Delta (l+\frac{1}{2}) \right]$
and $a_{\rm HCP}\left[
(j+\frac{1}{2}),\sqrt{3}(k+\frac{1}{6}),\Delta (l+\frac{1}{2}) \right]$
with integer $j,k,l$.
The energy per particle of HCP lattice for $\Re(s)>3$ thus reads as
\begin{eqnarray}
\zeta_{\mathsf{A 3}}(s) & = & \frac{1}{2^{s/6+1}} \Bigg\{
\sum_{(j,k,l) \ne (0,0,0)} \frac{1}{(j^2+3k^2+\frac{8}{3}l^2)^{s/2}} \nonumber\\ 
& & + \sum_{j,k,l}\frac{1}{[(j+1/2)^2+3(k+1/2)^2+\frac{8}{3}l^2]^{s/2}}\nonumber\\
& & + \sum_{j,k,l}\frac{1}{[j^2+3(k+2/3)^2+\frac{8}{3}(l+1/2)^2]^{s/2}}
\nonumber \\
& & + \sum_{j,k,l}\frac{1}{[(j+1/2)^2+3(k+1/6)^2+\frac{8}{3}(l+1/2)^2]^{s/2}}
\Bigg\} . \label{ehcp}
\end{eqnarray}

To obtain an analytic continuation of the energy to the whole complex plane
$s$, except for the point $s=3$, all sums over integers $j$ and $l$
can be transformed directly to Jacobi elliptic functions by using
the above explained summation techniques.
The transformation of the sums over integers $k$ with the shifts of $k$ by
$1/6$ and $2/3$ is not so straightforward, but still they are expressible
as series of products of Jacobi elliptic functions and the cos-functions. 
In particular, one gets
\begin{eqnarray} 
& & \frac{2^{s/6+1}\Gamma(s/2)}{\pi^{s/2}} \zeta_{\rm A3}(s) = 
\frac{4}{\sqrt{2}(s-3)}-\frac{2}{s} \nonumber \\
& & + \int_{0}^1 {\rm d} t\, t^{s/2-1} \left[
\theta_3({\rm e}^{-\pi t}) \theta_3({\rm e}^{-3\pi t})
\theta_3({\rm e}^{-8\pi t/3}) - \frac{1}{\sqrt{8}t^{3/2}} \right] \nonumber\\
& & + \int_{0}^1 {\rm d} t\, t^{(1-s)/2}
\left[ \frac{1}{\sqrt{8}} \theta_3({\rm e}^{-\pi t}) \theta_3({\rm e}^{-\pi t/3})
\theta_3({\rm e}^{-3\pi t/8}) - \frac{1}{t^{3/2}} \right] \nonumber\\
& & + \int_{0}^1 {\rm d} t \, t^{s/2-1} \left[
\theta_2({\rm e}^{-\pi t}) \theta_2({\rm e}^{-3\pi t})
\theta_3({\rm e}^{-8\pi t/3}) - \frac{1}{\sqrt{8}t^{3/2}} \right] \nonumber\\
& & + \frac{1}{\sqrt{8}} \int_{0}^1 {\rm d} t\, t^{(1-s)/2}
\theta_4({\rm e}^{-\pi t}) \theta_4({\rm e}^{-\pi t/3}) \theta_3({\rm e}^{-3\pi t/8})
\nonumber\\ 
& & + \int_{0}^1 {\rm d} t\, \frac{t^{s/2-1}}{\sqrt{3t}}
\left[ \theta_2({\rm e}^{-\pi t}) \theta_2({\rm e}^{-8\pi t/3}) 
-\frac{1}{\sqrt{8/3} t} \right] \nonumber\\
& & + 2 \int_{0}^1 {\rm d} t\,  \frac{t^{s/2-1}}{\sqrt{3t}}
\theta_2({\rm e}^{-\pi t}) \theta_2({\rm e}^{-8\pi t/3})
\sum_{j=1}^{\infty} \cos\left( \frac{\pi j}{3} \right)
{\rm e}^{-\pi j^2/(3 t)} \nonumber\\
& & + \int_{0}^1 {\rm d} t\, \frac{t^{(1-s)/2}}{\sqrt{8}}
\theta_4({\rm e}^{-\pi t}) \theta_4({\rm e}^{-8\pi t/3})
\left[ 1+2\sum_{j=1}^{\infty} \cos \left( \frac{\pi j}{3}\right)
{\rm e}^{-\pi tj^2/3} \right] \nonumber\\
& & + \int_{0}^1 {\rm d} t\,  \frac{t^{s/2-1}}{\sqrt{3t}}
\left[ \theta_3({\rm e}^{-\pi t}) \theta_2({\rm e}^{-8\pi t/3}) 
-\frac{1}{\sqrt{8/3} t} \right] \nonumber\\
& & + 2 \int_{0}^1 {\rm d} t\,  \frac{t^{s/2-1}}{\sqrt{3t}}
\theta_3({\rm e}^{-\pi t}) \theta_2({\rm e}^{-8\pi t/3})
\sum_{j=1}^{\infty} \cos\left( \frac{4\pi j}{3}\right)
{\rm e}^{-\pi j^2/(3 t)} \nonumber\\
& & + \int_{0}^1 {\rm d} t\, \frac{t^{(1-s)/2}}{\sqrt{8}}
\theta_3({\rm e}^{-\pi t}) \theta_4({\rm e}^{-3\pi t/8})
\left[ 1 + 2\sum_{j=1}^{\infty} \cos\left( \frac{4\pi j}{3}\right)
{\rm e}^{-\pi t j^2/3} \right] . \nonumber\\ & & \label{ehcpuni}
\end{eqnarray}
Here, in the numerical computation of infinite sums over cosine functions
one has to truncate the sums as follows $\sum_{j=1}^K$ where the upper cut
$K=1,2,3,\ldots$ is a positive integer.
It turns out that by increasing $K$ the convergence of the sums to the exact
value $(K\to\infty)$ is extremely quick. In particular, by using the symbolic computer language {\it Mathematica} it was checked for any real value of $s$ that the $K=4$ energy is determined with precision of 12 decimal digits, for $K=5$ the precision is increased to 16 decimal digits and so on.

\section*{Acknowledgments}

LB was supported by the Austrian Science Fund (FWF) and the German Research Foundation (DFG) through the joint project FR 4083/3-1/I 4354 during his stay in Vienna. L\v{S} and IT received support from VEGA Grant No. 2/0092/21.

\small{}

\end{document}